\documentclass[prb,twocolumn]{revtex4}
\usepackage{amssymb}
\usepackage{graphicx}

\begin{document}
\title{Temperature and fluence dependence of ultrafast phase separation dynamics in Pr$_{0.6}$Ca$_{0.4}$MnO$_3$ thin films}
\author{T. Mertelj$^{1,2}$, R. Yusupov$^{1}$, A. Gradi\v{s}ek$^{1}$, M. Filippi$^{3}$ , W. Prellier$^{3}$ and D. Mihailovic$^{1,2}$}
\date{\today}

\begin{abstract}
Temperature and fluence dependence of the transient photoinduced reflectivity and the magnetooptical Kerr angle was measured in two Pr$_{0.6}$Ca$_{0.4}$MnO$_3$ thin films subject to tensile and compressive substrate-induced strain. A photoinduced transient ferromagnetic  metallic (TFM) phase is found to form below $\sim$60K and $\sim$40K in the substrate-strained and substrate-compressed film, respectively. From the hysteresis loops a difference in the TFM cluster sizes and amount of photomodulation is observed at low temperatures and low excitation fluences in the films with different strain. Surprisingly, the characteristic timescale for the TFM phase photomodulation is virtually strain independent. At high excitation fluences, the cluster sizes and amount of photomodulation are independent on the substrate-induced strain.
\end{abstract}

\affiliation{$^{1}$Jozef Stefan Institute, Jamova 39, 1000
Ljubljana, Slovenia}

\affiliation{$^{2}$Faculty of Mathematics and Physics, Univ. of
Ljubljana, 1000 Ljubljana, Slovenia}

\affiliation{$^{3}$Laboratoire CRISMAT, CNRS UMR 6508, Bd du Marechal Juin, F-14050 Caen Cedex, France}


\maketitle

\section{Introduction}

Functional properties of the perovskite colossal-magnetoresistance  manganites are a result of the delicate balance between their insulating and metallic phases. The balance between these phases can be efficiently affected by external perturbations \cite{TomiokaAsamitsu1996,MiyanoTanaka1997,PrellierSimon2000,SaurelBrulet2006} which induce an insulator-metal (IM) transition. In (Pr,Ca)MnO$_3$ the photoinduced IM transition\cite{MiyanoTanaka1997} has attracted significant attention because it takes place on a sub-ps timescale and can be triggered by directly exciting the electronic\cite{FiebigMiyano2000,KidaTonouchi2001} or the lattice degrees of freedom\cite{RiniTobey2007}. 

In the (Pr,Ca)MnO$_3$ system the static IM transition is due to percolation between ferromagnetic metallic (FM) clusters\cite{SaurelBrulet2006} which form within an insulating phase as a result of different static external perturbations.\cite{TomiokaAsamitsu1996,PrellierSimon2000,SaurelBrulet2006} The \emph{dynamic} formation of the FM phase upon photoexcitation, on an ultrafast timescale, is however not yet fully understood. There are some reports on time resolved magnetooptical Kerr effect (TRMOKE) measurements which indicate that the short range ferromagnetic order might be created simultaneously with the metallic state on the sub-ps timescale.\cite{MiyasakaNakamura2006,MatsubaraOkimoto2007} However, in these reports only one component of the complex Kerr angle was reported without a detailed magnetic field dependence inhibiting elimination of the nonmagnetic components from the response.
Recently\cite{MerteljYusupovEPL2009} we were able to disentangle different components contributing to TRMOKE by comparing the time evolutions of the photoinduced reflectivity with TRMOKE at two different probe-photon energies as functions of the external magnetic field in (Pr$_{0.6}$Ca$_{0.4}$)MnO$_3$ thin films. We found no clear evidence that the magnetization is modified on the sub-picosecond scale. Instead, our data indicate that \emph{the photoinduced dynamical phase transition} of the ferromagnetic insulating (FI) phase into a transient ferromagnetic metallic (TFM) phase occurs on a slower 10-ps timescale at 5K. The TFM  phase is formed in clusters, which are up to $\sim$500 spins large and show no signs of decay on a 1.5-ns timescale.

In this paper we present a detailed temperature and fluence dependence of the time-resolved  photoinduced reflectivity and TRMOKE in the same samples to further elucidate properties of the TFM phase.

\section{Experimental}
Thin films of Pr$_{0.6}$Ca$_{0.4}$MnO$_3$ (PCMO60) with the thickness of $\sim$3000\AA \ were grown on (001)-oriented
SrTiO$_3$ (STO) and LaAlO$_3$ (LAO) substrates as described elsewhere.\cite{PrellierSimon2000} STO induces tensile and LAO compressive strain in the film.\cite{NelsonHill2004}  Due to the strain PCMO60 thin films are ferromagnetic and insulating below $\sim$120K.\cite{NelsonHill2004,MerteljYusupovAPL2008} In PCMO60/STO static magnetization and MOKE measurements indicate that at 5K a metastable ferromagnetic metallic (FM) phase coexists with a stable ferromagnetic insulating (FI) phase  at the surface of the film in a magnetic field below 1.1 T already. Near the surface, within the penetration depth of our optical probe which is $\sim$50 nm, both films show out-of-plane hard-axis magnetic anisotropy as described in detail in ref. \onlinecite{MerteljYusupovAPL2008}.

A linearly polarized pump beam with the photon energy 1.55 eV, the pulse length 50 fs and repetition frequency 250 kHz was focused to a 250-$\mu$m diameter spot on the sample in a nearly perpendicular geometry. The fluence of the pump pulses was $\mathcal{F}=150 \mu$J/cm$^2$ unless noted otherwise. The weaker probe beam with the photon energy $\hbar \omega_\mathrm{probe}=$  1.55 eV or 3.1 eV and the diameter 220 $\mu$m was focused to the same spot with the polarization perpendicular to that of the pump. The reflected probe beam was analyzed by a Wollaston prism and a pair of balanced silicon PIN photodiodes using standard lockin techniques.

\begin{figure}[h]
  \begin{center}
  \includegraphics[angle=-90,width=0.47\textwidth]{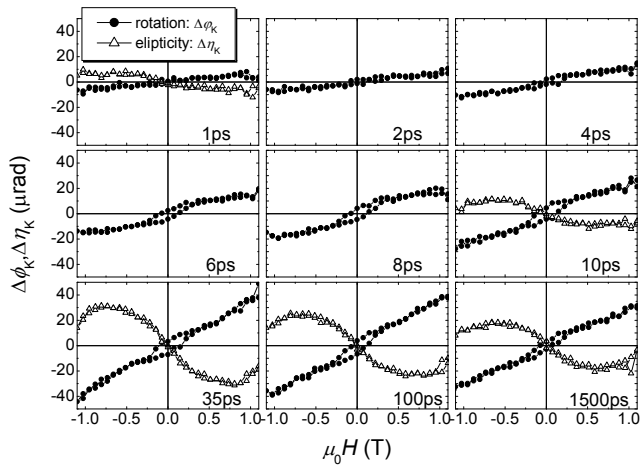}
  \end{center}
  \caption{The time evolution of $\Delta \phi_\mathrm{K}$ and  $\Delta \eta_\mathrm{K}$ $H$-loops in the PCMO60/STO sample at 5K and 1.55-eV pump photon energy.}
  \label{fig:time-evolution-both}
\end{figure}

Samples were mounted on a cold finger of an optical liquid-He flow cryostat equipped with CaF$_2$ windows placed in an 1.1-T electromagnet with hollow poles. All MOKE measurements were conducted in the polar geometry with the magnetic field perpendicular to the film. During TRMOKE measurements the static Kerr rotation $\phi_\mathrm{K} = Re\left(\Theta _\mathrm{K}\right)$ and Faraday rotation due to the cryostat window were compensated by a computer controlled rotation stage to keep the detector balanced at any magnetic field value.
The photoinduced complex Kerr angle transients $\Delta \Theta _\mathrm{K}=\frac{1}{2}\left[ \Delta \Theta \left( H\right) -\Delta \Theta \left( -H\right) \right] $  were obtained by subtracting photoinduced probe polarization change $\Delta \Theta$ taken at two opposite directions of the magnetic field to remove nonmagnetic contributions.
For all optical measurements the samples were first zero-field cooled (ZFC) to the lowest temperature. Subsequently data were collected at fixed temperatures during the warming part of a cycle.

\section{Results}

\subsection{Temperature dependence}

In Fig. \ref{fig:time-evolution-both} we plot the time evolution of the magnetic-field loops ($H$-loops) of the photoinduced Kerr rotation, $\Delta\phi _\mathrm{K} = \mathrm{Re}(\Delta\Theta _\mathrm{K})$, and the Kerr ellipticity, $\Delta\eta _\mathrm{K} = \mathrm{Im}(\Delta\Theta _\mathrm{K})$, at 5K and 1.55-eV probe-photon energy (PPE) in the PCMO60/STO sample. At 1 ps $\Delta\eta_\mathrm{K}$ shows saturation with increasing magnitude of the magnetic field $H$ and is positive with respect to the static\cite{MerteljYusupovAPL2008,MerteljYusupovEPL2009} $\eta _\mathrm{K}$ while $\Delta\phi _\mathrm{K}$ is linear with $H$ and negative with respect to the static\cite{MerteljYusupovAPL2008,MerteljYusupovEPL2009} $\phi _\mathrm{K}$. With increasing delay the magnitude of the response increases and shapes of the $H$-loops change. We observe a large qualitative difference between the shapes of $\Delta\eta _\mathrm{K}$ and $\Delta\phi _\mathrm{K}$  $H$-loops. While the $\Delta\phi _\mathrm{K}$  $H$-loops show hysteresis on top of an almost linear background the $\Delta\eta _\mathrm{K}$ $H$-loops show an undulation-like nonmonotonous magnetic field dependence.  Beyond the delay of $\sim$35 ps the shapes of the $H$-loops remain almost stationary while their magnitude slowly decays with increasing delay.

\begin{figure}[h]
  \begin{center}
  \includegraphics[angle=0,width=0.26\textwidth]{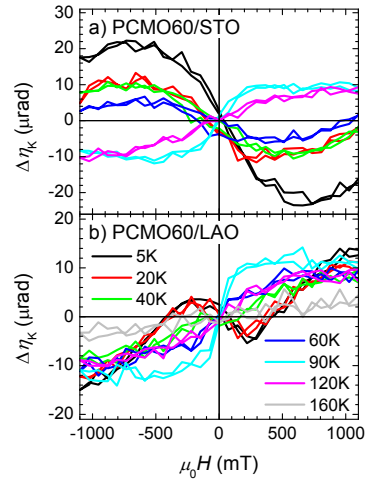}
  \end{center}
  \caption{Temperature dependence of $\Delta \eta_\mathrm{K}$ $H$-loops at 1500-ps delay and 1.55-eV probe-photon energy.}
  \label{fig:T-dep-eta}
\end{figure}

Previously we have shown that the difference between $\Delta\phi _\mathrm{K}$ and $\Delta\eta_\mathrm{K}$ $H$-loops is due to the coexistence of different magnetic phases with their fractions being modulated by the photoexcitation.\cite{MerteljYusupovEPL2009} The undulation in $\Delta\eta _\mathrm{K}$ $H$-loops at 1.55-eV and $\Delta\phi _\mathrm{K}$ $H$-loops at 3.1-eV  was associated with the TFM phase. In Fig. \ref{fig:T-dep-eta} we hence show the temperature dependence of $\Delta\eta _\mathrm{K}$ at 1.55-eV in both samples. 
With increasing $T$ the amplitude of the undulation, which is weaker in the PCMO60/LAO sample, decreases and the shoulders vanish above 40K in PCMO60/LAO and 60K in PCMO60/STO indicating absence of the TFM phase. At 90K and above the $H$-loops are virtually identical in both samples and qualitatively different than at lower temperatures. 

\begin{figure}[hbt]
  \begin{center}
  \includegraphics[angle=-90,width=0.40\textwidth]{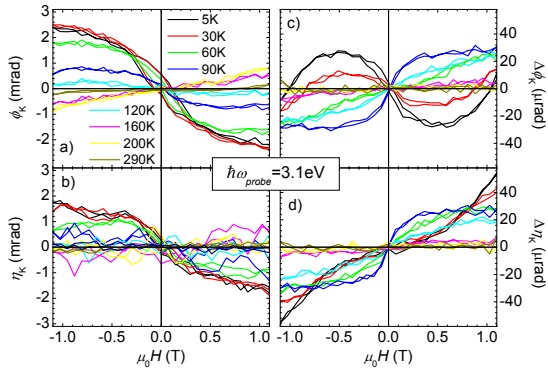}
  \end{center}
  \caption{Temperature dependence of the static $\phi_\mathrm{K}$ (a) and  $\eta_\mathrm{K}$ (b) $H$-loops in PCMO60/STO  compared to the photoinduced counterparts $\Delta \phi_\mathrm{K}$ (c) and  $\Delta \eta_\mathrm{K}$ (d) at 1500-ps delay and 3.1-eV probe-photon energy. }
    \label{fig:T-dep-all-H}
\end{figure}

To compare $T$-dependence of the photoinduced $\Delta\phi _\mathrm{K}$ ($\Delta\eta _\mathrm{K}$) with the static $\phi _\mathrm{K}$ ($\eta _\mathrm{K}$) we choose 3.1-eV PPE to enable reliable simultaneous observation of both static components.\footnote{At 3.1-eV PPE, contrary to 1.55-eV PPE, both $\phi _\mathrm{K}$ and $\eta _\mathrm{K}$ have magnitudes above the systematic error.\cite{MerteljYusupovAPL2008} However, the contributions of different magnetic phases to the dynamic responses are not so well separated at 3.1-eV PPE as at 1.55-eV PPE.\cite{MerteljYusupovEPL2009}} At this PPE $\Delta\phi _\mathrm{K}$ behaves similarly to $\Delta\eta _\mathrm{K}$ at 1.55-eV PPE and $\Delta\eta _\mathrm{K}$ similarly  to $\Delta\phi _\mathrm{K}$ at 1.55-eV PPE. Again we observe a qualitative change in the shape of the dynamic $H$-loops at $\sim$60K (see Figs. \ref{fig:T-dep-all-H}c, \ref{fig:T-dep-all-H}d). Above this temperature the shapes of $\Delta\phi _\mathrm{K}$ and $\Delta\eta_\mathrm{K}$ $H$-loops become virtually identical. There is no corresponding large qualitative change of the shapes of the static $H$-loops around 60K (see Figs. \ref{fig:T-dep-all-H}a, \ref{fig:T-dep-all-H}b) apart from disappearance of the hysteresis in the $\phi _\mathrm{K}$ $H$-loops above 60K.

\begin{figure}[h]
  \begin{center}
  \includegraphics[angle=0,width=0.40\textwidth]{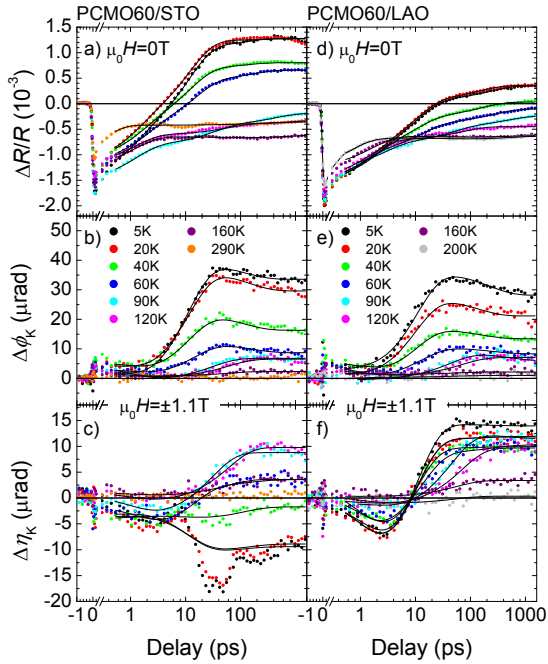}
  \end{center}
  \caption{Temperature dependence of $\Delta R/R$, $\Delta \phi_\mathrm{K}$ and  $\Delta \eta_\mathrm{K}$  transients at 1.55-eV probe-photon energy for the PCMO60/STO sample a), b), c) and the PCMO60/LAO sample d), e), f). The thin lines represent fits discussed in text.}
    \label{fig:T-dep-all-dly}
\end{figure}

The qualitative change around  60K is observed also in the delay dependence of $\Delta \eta_\mathrm{K}$ at 1.55-eV PPE in the PCMO60/STO sample (see. Fig. \ref{fig:T-dep-all-dly}c). Below $\sim$ 60K $\Delta \eta_\mathrm{K}$ at 1.1 T is negative at all positive delays due to the presence of the undulation, while above $\sim$ 60K a zero crossing at $\sim$20ps is observed, which is present at all $T$ below 160K in the PCMO60/LAO sample, albeit at somewhat shorter delay. In addition to the different sign $\Delta \eta_\mathrm{K}$-traces in the PCMO60/STO sample also show presence of a strongly-damped oscillatory component at low temperatures, which is not observed in the PCMO60/LAO sample. Contrary to $\Delta \eta_\mathrm{K}$, the delay dependence of $\Delta \phi_\mathrm{K}$ is virtually identical in both samples at all $T$ (see. Figures \ref{fig:T-dep-all-dly} b) and \ref{fig:T-dep-all-dly} e)).

\begin{figure}[h]
  \begin{center}
  \includegraphics[angle=0,width=0.40\textwidth]{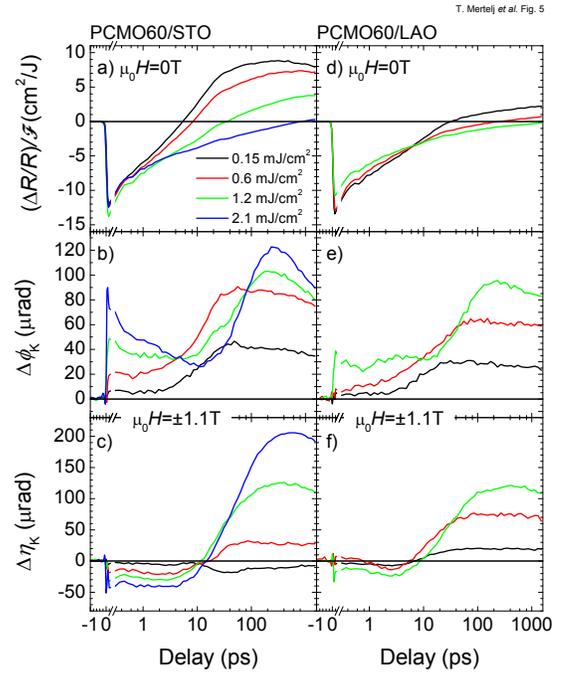}
  \end{center}
  \caption{Fluence dependence of $\Delta R/R$, $\Delta \phi_\mathrm{K}$ and  $\Delta \eta_\mathrm{K}$  transients at 5K and 1.55-eV probe-photon energy for the PCMO60/STO sample a), b), c) and the PCMO60/LAO sample d), e), f). Note that $\Delta R/R$ is normalized with the fluence while TRMOKE transients are not. }
    \label{fig:P-dep-dly}
\end{figure}

\subsection{Excitation fluence dependence}

Next we turn to the excitation fluence ($\mathcal{F}$) dependence of the transients at 5K. The sub-ps $\Delta R/R$ shows almost linear dependence on $\mathcal{F}$ (see Fig. \ref{fig:P-dep-dly} a), d)) in both samples. On timescales longer than $\sim$0.5 ps however, the response becomes slower and the magnitude saturates with increasing $\mathcal{F}$. The saturation is less pronounced in the PCMO60/LAO sample. The $\Delta \eta_\mathrm{K}$ transients shown in Fig. \ref{fig:P-dep-dly} c), f) show saturation with $\mathcal{F}$ on all timescales while the saturation is less pronounced for $\Delta \phi_\mathrm{K}$ on the sub-ps timescale.\footnote{We can not exclude some contamination of $\Delta \phi_\mathrm{K}$ transient with $\Delta R/R$ transient at the highest $\mathcal{F}$ due to an unperfect detector balancing.} As for the $\Delta R/R$ transients, the risetime of the slower part of the response increases $\sim$10 times when $\mathcal{F}$ is increased from 0.15 mJ/cm$^2$ to 2.1 mJ/cm$^2$ and 1.2 mJ/cm$^2$ in the PCMO60/STO and PCMO60/LAO sample, respectively.

\begin{figure}[h]
  \begin{center}
  \includegraphics[angle=-90,width=0.40\textwidth]{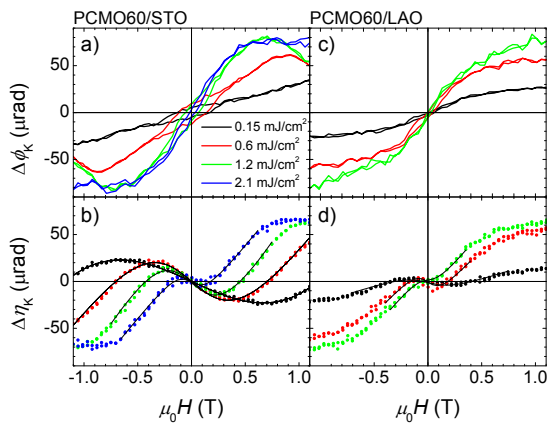}
  \end{center}
  \caption{Fluence dependence of $\Delta \phi_\mathrm{K}$ and  $\Delta \eta_\mathrm{K}$ $H$-loops at 35-ps delay, temperature 5K and 1.55-eV probe-photon energy for the PCMO60/STO sample a), b) and the PCMO60/LAO sample c), d). The thin lines in panels b) and d) represent fits discussed in the text.}
    \label{fig:P-dep-H}
\end{figure}

The sign change of the $\Delta \eta_\mathrm{K}$-transients in the PCMO60/STO sample with increasing $\mathcal{F}$ originates, similarly as for the $T$-dependence, in vanishing contribution of the component associated with the TFM-phase-fraction modulation (see Fig. \ref{fig:P-dep-H}). The decrease of the magnitude of the undulation in $\Delta\eta_\mathrm{K}$ is accompanied the shift of extrema towards a lower magnetic field. At higher fields above $\sim$0.8 T a saturation appears in both $\Delta\eta_\mathrm{K}$ and $\Delta\phi_\mathrm{K}$ $H$-loops in the PCMO60/STO sample at the highest fluences. As a result of the saturation the shapes of the corresponding $H$-loops become almost identical when one compares PCMO60/STO and PCMO60/LAO samples at the highest fluences. At the highest $\mathcal{F}$ signs of a weak hysteresis appear in $\Delta\eta_\mathrm{K}$ $H$-loops in the region of undulation in both samples.

\section{Discussion}

In our previous work\cite{MerteljYusupovEPL2009} we have shown that the origin of the complex behavior of TRMOKE at 5K in PCMO60 thin films is due to the transient photo-modulation of the fractions of different magnetic phases. Upon photoexcitation the amount of the FI phase is decreased $\sim$3\% at $\mathcal{F}=0.15$mJ/cm$^2$ and at least partly converted into the TFM phase. The increase of the TFM-phase fraction was associated with the undulation in $H$-loops of $\Delta\eta_\mathrm{K}$ at 1.55-eV PPE and $\Delta\phi_\mathrm{K}$ at 3.1-eV PPE, which appear on top of the response originating in the decrease of the FI-phase fraction. The TFM phase is created in the form of clusters with sizes of several hundred of electron spins as determined from the Brillouin function fits. 
In the PCMO60/STO sample in addition to the FI and TFM phases also a static FM (SFM) phase was detected. This phase coexists with the FI phase also in the absence of the photoexcitation\cite{MerteljYusupovAPL2008} and is responsible for the presence of the hysteretic part of the static and low-$\mathcal{F}$ dynamic responses in the PCMO60/STO sample.

To estimate the dependence of the size of  the clusters on $\mathcal{F}$ we assume, that at 1.55-eV PPE\footnote{We choose 1.55-eV PPE since it is free from the hysteretic contribution from the SFM phase.}  the background contribution due to the other phases in $\Delta\eta_\mathrm{K}$ is approximately a linear function of $H$ in the region of undulation, and fit a sum of a linear and Brillouin function to $\Delta\eta_\mathrm{K}$ (see Fig. \ref{fig:P-dep-H}) to extract the cluster sizes. The sizes obtained from the fits are summarized in Table \ref{tab:csizes}. A monotonous increase of the cluster sizes with increasing $\mathcal{F}$ is observed in the PCMO60/STO sample in the complete  $\mathcal{F}$-range while in the PCMO60/LAO sample the cluster size saturates above 0.6 mJ/cm$^2$. Together with increase of the cluster size we observe a decrease of the magnitude of the undulation. 

To understand this behavior we should take into account that in our experiment the pump pulses arrive each 4 $\mu$s. If the decay of the TFM phase is slower than 4 $\mu$s its fraction builds up. When the steady state conditions are achieved, such as in our experiment, the undulation in $\Delta\eta_\mathrm{K}$ represents just an incremental increase of the fraction which compensates its decay between the successive pump pulses. The decrease of the magnitude of the undulation therefore indicates that the TFM phase becomes quasi-static and that its amount saturates at high fluences with simultaneous slowing-down of the dynamics. Such scenario is supported also by appearance of the weak hysteresis in $\Delta\eta_\mathrm{K}$ at the highest $\mathcal{F}$ which indicates a long term memory for the magnetization orientation of the TFM phase. The weak hysteresis also suggests that at the highest $\mathcal{F}$ the single-domain cluster model used to obtain cluster size from the fits probably breaks down. The apparent saturation of the cluster size in the PCMO60/LAO sample at high $\mathcal{F}$ can therefore be linked with the emergence of multi-domain quasi-static TFM clusters. 

Let us now analyze the temperature dependence of the TFM phase. Contrary to the case of increasing $\mathcal{F}$ there is no evidence for a change of the cluster size when the temperature is increased and the amount of the TFM phase decreases. At 90K and above the shapes of the dynamic $H$-loops are identical in both samples for both components of the Kerr angle. This observation is consistent with the presence of a single magnetic phase in this $T$-range corresponding to the majority FI phase. The relative sign  of the dynamic $H$-loops with respect to the static ones corresponds to decrease of the magnetization. We believe that the main contribution to the demagnetization is the decrease of the FI-phase fraction.

\begin{table}[t]
\centering
\begin{tabular}{@{\extracolsep{\fill}}cr@{.}l|cr@{$\pm$}l|cr@{$\pm$}l}
\multicolumn{3}{c|}{}&\multicolumn{3}{c|}{PCMO60/STO} &\multicolumn{3}{c}{PCMO60/LAO}\\
\hline
\multicolumn{3}{c|}{$\mathcal{F}$ (mJ/cm$^2$)}&  \multicolumn{6}{c}{cluster size}\\
\hline
&0&15   && 50&10    && 400&100 \\
&0&6    && 160&10   && 700&200 \\
&1&2    && 260&20   && 700&300\\
&2&1    && 600&100  &\multicolumn{2}{c}{-}\\
\end{tabular}
\caption{The cluster sizes obtained from the fits in Fig. \ref{fig:P-dep-H}. For details see text.}
\label{tab:csizes}
\end{table}

\begin{figure}
  \begin{center}
  \includegraphics[angle=0,width=0.25\textwidth]{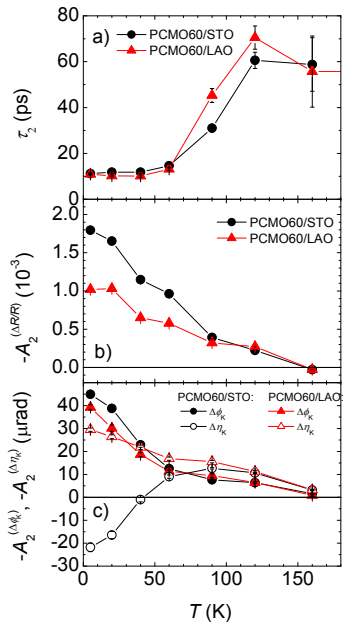}
  \end{center}
  \caption{Temperature dependence of the phase-fraction-photomodulation component  risetime a) and amplitudes b), c) at 1.55-eV probe-photon energy.}
    \label{fig:fit-10ps-T}
\end{figure}

We can fit $\Delta \phi_\mathrm{K}$, $\Delta \eta_\mathrm{K}$ and $\Delta R/R$ delay scans at any given $T$ in a given sample for the delay longer than 0.5 ps  with a sum of three exponential functions: $y(t)=A_{0}+\sum A_{i}\exp \left( -t/\tau _{i}\right)$, where the relaxation times, $\tau _{1}\approx 1$ps, $\tau _{2}\approx 10$ps and $\tau _{3}\approx 200$ps \emph{are shared among all the traces}, and the amplitudes $A_0$ and $A_i$ are independent parameters for each trace separately. The only exceptions are $\Delta\eta _\mathrm{K}$-traces below 40K in the PCMO60/STO sample where the strongly-damped oscillatory component is observed. The temperature dependence of the fit parameters for the most prominent 10-ps component, which we associate with the photomodulation of the phase fractions\cite{MerteljYusupovEPL2009}, are shown in Fig. \ref{fig:fit-10ps-T}. 

The phase-fraction photomodulation (PFP) rise time ($\tau_2$) has almost identical value and shows very similar $T$-dependence in both samples. It is almost constant up to 60K and starts to increase concurrently with the disappearance of the TFM phase above 40K in PCMO60/LAO and 60K in PCMO60/STO showing a divergence at the Curie temperature. At low $T$ the amplitudes of the PFP-component in $\Delta R/R$-transients ($A_2^{(\Delta R/R)}$) in PCMO60/STO are consistently higher than in the PCMO60/LAO sample and the presence of the undulation due to the TFM phase is reflected in different $T$-dependencies of $A_2^{(\Delta \phi_\mathrm{K})}$ with respect to $A_2^{(\Delta \eta_\mathrm{K})}$. While $A_2^{(\Delta \phi_\mathrm{K})}$ behaves similarly to $A_2^{(\Delta R/R)}$ with increasing $T$, $A_2^{(\Delta \eta_\mathrm{K})}$ changes sign in the PMCO60/STO sample and shows less steep $T$-dependence in the  PCMO60/LAO sample.

At 90K and above all the dynamical responses critically slow down and are consistent with the photoinduced decrease of the single FI-phase fraction since the $T$-dependence of $A_2^{(\Delta R/R)}$ closely follows that of $A_2^{(\Delta \phi_\mathrm{K})}$ and $A_2^{(\Delta \eta_\mathrm{K})}$ in this $T$-range.

\section{Conclusions}

We presented a detailed fluence and temperature dependence study of the photoinduced TFM phase fraction modulation in Pr$_{0.6}$Ca$_{0.4}$MnO$_3$ thin films on two different substrates with different substrate induced strain. The TFM phase is photoinduced from the majority FI phase below $\sim$60K in the tensile-strained PCMO60/STO film and below $\sim$40K in the compressive-strained PCMO60/LAO film. 

At low temperature and fluence the tensile induced strain in the PCMO60/STO film results in smaller TFM-phase clusters and larger magnitude of the TFM-phase fraction photomodulation than the compressive induced strain in the PCMO60/LAO film. There is no difference in the characteristic modulation timescale of 10 ps between the two films. 

At high fluences, where the clusters become larger and multidomain, a saturation of the amount of the TFM-phase modulation, concurrent with slowing down of the dynamics, is observed. In addition, irrespectively of the strain, both films show very similar cluster size and amount of modulation. 

There is also virtually no difference between the behavior in both films above 90K where only the ferromagnetic insulating phase is present. In this $T$-range we observe a decrease of the phase fraction of the majority FI phase upon photoexcitation and critical slowing down of the timescale when approaching the Curie temperature.

\begin{acknowledgments}
We acknowledge fruitful discussions with V.V. Kabanov.

\end{acknowledgments}

\bibliography{biblio}

\begin{thebibliography}{12}
\expandafter\ifx\csname natexlab\endcsname\relax\def\natexlab#1{#1}\fi
\expandafter\ifx\csname bibnamefont\endcsname\relax
  \def\bibnamefont#1{#1}\fi
\expandafter\ifx\csname bibfnamefont\endcsname\relax
  \def\bibfnamefont#1{#1}\fi
\expandafter\ifx\csname citenamefont\endcsname\relax
  \def\citenamefont#1{#1}\fi
\expandafter\ifx\csname url\endcsname\relax
  \def\url#1{\texttt{#1}}\fi
\expandafter\ifx\csname urlprefix\endcsname\relax\def\urlprefix{URL }\fi
\providecommand{\bibinfo}[2]{#2}
\providecommand{\eprint}[2][]{\url{#2}}

\bibitem[{\citenamefont{Tomioka et~al.}(1996)\citenamefont{Tomioka, Asamitsu,
  Kuwahara, Moritomo, and Tokura}}]{TomiokaAsamitsu1996}
\bibinfo{author}{\bibfnamefont{Y.}~\bibnamefont{Tomioka}},
  \bibinfo{author}{\bibfnamefont{A.}~\bibnamefont{Asamitsu}},
  \bibinfo{author}{\bibfnamefont{H.}~\bibnamefont{Kuwahara}},
  \bibinfo{author}{\bibfnamefont{Y.}~\bibnamefont{Moritomo}}, \bibnamefont{and}
  \bibinfo{author}{\bibfnamefont{Y.}~\bibnamefont{Tokura}},
  \bibinfo{journal}{Phys. Rev. {\bf B}} \textbf{\bibinfo{volume}{53}},
  \bibinfo{pages}{R1689} (\bibinfo{year}{1996}).

\bibitem[{\citenamefont{{Miyano} et~al.}(1997)\citenamefont{{Miyano}, {Tanaka},
  {Tomioka}, and {Tokura}}}]{MiyanoTanaka1997}
\bibinfo{author}{\bibfnamefont{K.}~\bibnamefont{{Miyano}}},
  \bibinfo{author}{\bibfnamefont{T.}~\bibnamefont{{Tanaka}}},
  \bibinfo{author}{\bibfnamefont{Y.}~\bibnamefont{{Tomioka}}},
  \bibnamefont{and} \bibinfo{author}{\bibfnamefont{Y.}~\bibnamefont{{Tokura}}},
  \bibinfo{journal}{Phys. Rev. Lett.} \textbf{\bibinfo{volume}{78}},
  \bibinfo{pages}{4257} (\bibinfo{year}{1997}).

\bibitem[{\citenamefont{Prellier et~al.}(2000)\citenamefont{Prellier, Simon,
  Haghiri-Gosnet, Mercey, and Raveau}}]{PrellierSimon2000}
\bibinfo{author}{\bibfnamefont{W.}~\bibnamefont{Prellier}},
  \bibinfo{author}{\bibfnamefont{C.}~\bibnamefont{Simon}},
  \bibinfo{author}{\bibfnamefont{A.~M.} \bibnamefont{Haghiri-Gosnet}},
  \bibinfo{author}{\bibfnamefont{B.}~\bibnamefont{Mercey}}, \bibnamefont{and}
  \bibinfo{author}{\bibfnamefont{B.}~\bibnamefont{Raveau}},
  \bibinfo{journal}{Phys. Rev. {\bf B}} \textbf{\bibinfo{volume}{62}},
  \bibinfo{pages}{R16337} (\bibinfo{year}{2000}).

\bibitem[{\citenamefont{{Saurel} et~al.}(2006)\citenamefont{{Saurel},
  {Br{\^u}let}, {Heinemann}, {Martin}, {Mercone}, and
  {Simon}}}]{SaurelBrulet2006}
\bibinfo{author}{\bibfnamefont{D.}~\bibnamefont{{Saurel}}},
  \bibinfo{author}{\bibfnamefont{A.}~\bibnamefont{{Br{\^u}let}}},
  \bibinfo{author}{\bibfnamefont{A.}~\bibnamefont{{Heinemann}}},
  \bibinfo{author}{\bibfnamefont{C.}~\bibnamefont{{Martin}}},
  \bibinfo{author}{\bibfnamefont{S.}~\bibnamefont{{Mercone}}},
  \bibnamefont{and} \bibinfo{author}{\bibfnamefont{C.}~\bibnamefont{{Simon}}},
  \bibinfo{journal}{Phys. Rev. {\bf B}} \textbf{\bibinfo{volume}{73}},
  \bibinfo{pages}{094438} (\bibinfo{year}{2006}).

\bibitem[{\citenamefont{{Fiebig} et~al.}(2000)\citenamefont{{Fiebig}, {Miyano},
  {Tomioka}, and {Tokura}}}]{FiebigMiyano2000}
\bibinfo{author}{\bibfnamefont{M.}~\bibnamefont{{Fiebig}}},
  \bibinfo{author}{\bibfnamefont{K.}~\bibnamefont{{Miyano}}},
  \bibinfo{author}{\bibfnamefont{Y.}~\bibnamefont{{Tomioka}}},
  \bibnamefont{and} \bibinfo{author}{\bibfnamefont{Y.}~\bibnamefont{{Tokura}}},
  \bibinfo{journal}{Applied Physics B: Lasers and Optics}
  \textbf{\bibinfo{volume}{71}}, \bibinfo{pages}{211} (\bibinfo{year}{2000}).

\bibitem[{\citenamefont{{Kida} and {Tonouchi}}(2001)}]{KidaTonouchi2001}
\bibinfo{author}{\bibfnamefont{N.}~\bibnamefont{{Kida}}} \bibnamefont{and}
  \bibinfo{author}{\bibfnamefont{M.}~\bibnamefont{{Tonouchi}}},
  \bibinfo{journal}{Applied Physics Letters} \textbf{\bibinfo{volume}{78}},
  \bibinfo{pages}{4115} (\bibinfo{year}{2001}),
  \eprint{arXiv:cond-mat/0008298}.

\bibitem[{\citenamefont{{Rini} et~al.}(2007)\citenamefont{{Rini}, {Tobey},
  {Dean}, {Itatani}, {Tomioka}, {Tokura}, {Schoenlein}, and
  {Cavalleri}}}]{RiniTobey2007}
\bibinfo{author}{\bibfnamefont{M.}~\bibnamefont{{Rini}}},
  \bibinfo{author}{\bibfnamefont{R.}~\bibnamefont{{Tobey}}},
  \bibinfo{author}{\bibfnamefont{N.}~\bibnamefont{{Dean}}},
  \bibinfo{author}{\bibfnamefont{J.}~\bibnamefont{{Itatani}}},
  \bibinfo{author}{\bibfnamefont{Y.}~\bibnamefont{{Tomioka}}},
  \bibinfo{author}{\bibfnamefont{Y.}~\bibnamefont{{Tokura}}},
  \bibinfo{author}{\bibfnamefont{R.~W.} \bibnamefont{{Schoenlein}}},
  \bibnamefont{and}
  \bibinfo{author}{\bibfnamefont{A.}~\bibnamefont{{Cavalleri}}},
  \bibinfo{journal}{Nature} \textbf{\bibinfo{volume}{449}}, \bibinfo{pages}{72}
  (\bibinfo{year}{2007}).

\bibitem[{\citenamefont{{Miyasaka} et~al.}(2006)\citenamefont{{Miyasaka},
  {Nakamura}, {Ogimoto}, {Tamaru}, and {Miyano}}}]{MiyasakaNakamura2006}
\bibinfo{author}{\bibfnamefont{K.}~\bibnamefont{{Miyasaka}}},
  \bibinfo{author}{\bibfnamefont{M.}~\bibnamefont{{Nakamura}}},
  \bibinfo{author}{\bibfnamefont{Y.}~\bibnamefont{{Ogimoto}}},
  \bibinfo{author}{\bibfnamefont{H.}~\bibnamefont{{Tamaru}}}, \bibnamefont{and}
  \bibinfo{author}{\bibfnamefont{K.}~\bibnamefont{{Miyano}}},
  \bibinfo{journal}{Phys. Rev. {\bf B}} \textbf{\bibinfo{volume}{74}},
  \bibinfo{pages}{012401} (\bibinfo{year}{2006}).

\bibitem[{\citenamefont{{Matsubara} et~al.}(2007)\citenamefont{{Matsubara},
  {Okimoto}, {Ogasawara}, {Tomioka}, {Okamoto}, and
  {Tokura}}}]{MatsubaraOkimoto2007}
\bibinfo{author}{\bibfnamefont{M.}~\bibnamefont{{Matsubara}}},
  \bibinfo{author}{\bibfnamefont{Y.}~\bibnamefont{{Okimoto}}},
  \bibinfo{author}{\bibfnamefont{T.}~\bibnamefont{{Ogasawara}}},
  \bibinfo{author}{\bibfnamefont{Y.}~\bibnamefont{{Tomioka}}},
  \bibinfo{author}{\bibfnamefont{H.}~\bibnamefont{{Okamoto}}},
  \bibnamefont{and} \bibinfo{author}{\bibfnamefont{Y.}~\bibnamefont{{Tokura}}},
  \bibinfo{journal}{Phys. Rev. Lett.} \textbf{\bibinfo{volume}{99}},
  \bibinfo{pages}{207401} (\bibinfo{year}{2007}).

\bibitem[{\citenamefont{Mertelj et~al.}(2009)\citenamefont{Mertelj, Yusupov,
  Gradi{\v{s}}ek, Filippi, Prellier, and Mihailovic}}]{MerteljYusupovEPL2009}
\bibinfo{author}{\bibfnamefont{T.}~\bibnamefont{Mertelj}},
  \bibinfo{author}{\bibfnamefont{R.}~\bibnamefont{Yusupov}},
  \bibinfo{author}{\bibfnamefont{A.}~\bibnamefont{Gradi{\v{s}}ek}},
  \bibinfo{author}{\bibfnamefont{M.}~\bibnamefont{Filippi}},
  \bibinfo{author}{\bibfnamefont{W.}~\bibnamefont{Prellier}}, \bibnamefont{and}
  \bibinfo{author}{\bibfnamefont{D.}~\bibnamefont{Mihailovic}},
  \bibinfo{journal}{Europhysics Letters} \textbf{\bibinfo{volume}{86}},
  \bibinfo{pages}{57003} (\bibinfo{year}{2009}).

\bibitem[{\citenamefont{{Nelson} et~al.}(2004)\citenamefont{{Nelson}, {Hill},
  {Gibbs}, {Rajeswari}, {Biswas}, {Shinde}, {Greene}, {Venkatesan}, {Millis},
  {Yokaichiya} et~al.}}]{NelsonHill2004}
\bibinfo{author}{\bibfnamefont{C.~S.} \bibnamefont{{Nelson}}},
  \bibinfo{author}{\bibfnamefont{J.~P.} \bibnamefont{{Hill}}},
  \bibinfo{author}{\bibfnamefont{D.}~\bibnamefont{{Gibbs}}},
  \bibinfo{author}{\bibfnamefont{M.}~\bibnamefont{{Rajeswari}}},
  \bibinfo{author}{\bibfnamefont{A.}~\bibnamefont{{Biswas}}},
  \bibinfo{author}{\bibfnamefont{S.}~\bibnamefont{{Shinde}}},
  \bibinfo{author}{\bibfnamefont{R.~L.} \bibnamefont{{Greene}}},
  \bibinfo{author}{\bibfnamefont{T.}~\bibnamefont{{Venkatesan}}},
  \bibinfo{author}{\bibfnamefont{A.~J.} \bibnamefont{{Millis}}},
  \bibinfo{author}{\bibfnamefont{F.}~\bibnamefont{{Yokaichiya}}},
  \bibnamefont{et~al.}, \bibinfo{journal}{Journal of Physics Condensed Matter}
  \textbf{\bibinfo{volume}{16}}, \bibinfo{pages}{13} (\bibinfo{year}{2004}),
  \eprint{arXiv:cond-mat/0303228}.

\bibitem[{\citenamefont{Mertelj et~al.}(2008)\citenamefont{Mertelj, Yusupov,
  Filippi, Prellier, and Mihailovic}}]{MerteljYusupovAPL2008}
\bibinfo{author}{\bibfnamefont{T.}~\bibnamefont{Mertelj}},
  \bibinfo{author}{\bibfnamefont{R.}~\bibnamefont{Yusupov}},
  \bibinfo{author}{\bibfnamefont{M.}~\bibnamefont{Filippi}},
  \bibinfo{author}{\bibfnamefont{W.}~\bibnamefont{Prellier}}, \bibnamefont{and}
  \bibinfo{author}{\bibfnamefont{D.}~\bibnamefont{Mihailovic}},
  \bibinfo{journal}{Applied Physics Letters} \textbf{\bibinfo{volume}{93}},
  \bibinfo{pages}{042512} (\bibinfo{year}{2008}).

\end{thebibliography}

\end{document}